\documentclass{emulateapj}
\usepackage{graphicx}
\begin{document}

\def\etal{et al.\ \rm}
\def\ba{\begin{eqnarray}}
\def\ea{\end{eqnarray}}
\def\etal{et al.\ \rm}
\def\Fdw{F_{\rm dw}}
\def\Tex{T_{\rm ex}}
\def\Fdis{F_{\rm dw,dis}}
\def\Fnu{F_\nu}
\def\FJ{F_{J}}
\def\FJE{F_{J,{\rm Edd}}}

\title{Planet formation in small separation binaries: not 
so excited after all}

\author{Roman R. Rafikov\altaffilmark{1}}
\altaffiltext{1}{Department of Astrophysical Sciences, 
Princeton University, Ivy Lane, Princeton, NJ 08540; 
rrr@astro.princeton.edu}


\begin{abstract}
Existence of planets is binaries with relatively small 
separations (around 20 AU), such as $\alpha$ Centauri or 
$\gamma$ Cephei poses severe challenges to standard planet 
formation theories. The problem lies in the vigorous secular 
excitation of planetesimal eccentricities at separations of 
several AU, where some of the planets are found,
by the massive, eccentric stellar companions. High relative 
velocities of planetesimals preclude their growth in mutual 
collisions for a wide range of sizes, from below 1 km up to 
several hundred km, resulting in fragmentation barrier 
to planet formation. Here we show that rapid apsidal 
precession of planetesimal orbits, caused by the gravity of 
the circumstellar protoplanetary disk, acts to strongly
reduce eccentricity excitation, lowering planetesimal 
velocities by an order of magnitude or even 
more at 1 AU. By examining the details of planetesimal dynamics 
we demonstrate that this effect eliminates fragmentation barrier 
for in-situ growth of planetesimals as small as $\lesssim 10$ km 
even at separations as wide as 2.6 AU (semi-major axis of the 
giant planet in HD 196885), provided that the circumstellar 
protoplanetary disk is relatively massive, $\sim 0.1M_\odot$.
\end{abstract}



\section{Introduction.}  
\label{sect:intro}


About $20\%$ of planets detected via stellar radial velocity
variations reside in binaries (Desidera \& Barbieri 
2007). The majority of these systems are wide separation 
binaries, with semi-major axis $a_b\gtrsim 30$ AU. 
At the same time, four relatively small separation binaries 
with $a_b\approx 20$ AU (HD 196885, $\gamma$ Cephei, 
Gl 86 and HD 41004; Chauvin \etal 2011) are also known to 
harbor giant planets with projected masses 
$M_{pl}\sin i\approx (1.6-4)M_J$. In these systems the mass 
of the secondary star (we call ``secondary'' the 
binary component other than the star orbited by the planet, 
which we denote as ``primary'') $M_s$
is found to be close to $0.4M_\odot$ and binary eccentricity 
$e_b$ is close to $0.4$. Also, Dumusque \etal (2012)
have recently announced an Earth-mass companion to $\alpha$
Centauri B, a member of the binary (or, possibly, a triple)
with $a_b=17.6$ AU, $e_b=0.52$, and $M_s=1.1M_\odot$. 
In this system planet orbits the star at 
$\approx 0.04$ AU separation.

The uniqueness of these systems lies in the fact that forming 
planets in them is known to provide extreme challenge 
to planet formation theories (Zhou \etal 2012). With the 
exception of $\alpha$ Cen and Gl 86, planets in these 
binaries reside
in rather wide orbits, with planetary semi-major axes 
$a_{pl}\approx 1.6-2.6$ AU. In-situ formation of these 
gas giants is expected to proceed through continuous agglomeration  
of planetesimals at these locations, starting from very 
small objects (easily $\lesssim 1$ km). However, gravitational 
perturbations from the eccentric stellar companion inevitably 
result in rapid secular evolution (Heppenheimer 1978), driving 
planetesimal eccentricities far above the 
level at which bodies can avoid destruction in 
mutual collisions (Th\'ebault \etal 2008). This problem,
which is often called {\it collisional} or {\it fragmentation 
barrier}, is especially severe for small planetesimals, $1-10^2$ km 
in size, for which the ratio of binding to kinetic energy is 
small. It is also more pronounced far from the primary, where 
the secular forcing by the companion is strongest and 
planetesimal eccentricities are high.

Marzari \& Scholl (2000) suggested that a combination of secular 
forcing by the companion and gas drag acting on small 
($1-10$ km) planetesimals leads to apsidal alignment of their 
orbits, resulting in smaller relative velocities, and allowing 
colliding objects to grow. However, Th\'ebault \etal 
(2006, 2008) demonstrated that the planetesimal size-dependence 
of apsidal alignment acts to break orbital phasing between 
objects of different sizes, resulting in high velocity 
collisions between them and reinforcing collisional 
barrier.

Interestingly, most studies of planetesimal growth in 
small separation binaries have included the effect of 
the protoplanetary disk on planetesimal dynamics only via 
associated gas drag (Th\'ebault \etal 2004, 
2006, 2008, 2009; Paardekooper \etal 2008; Paardekooper \& 
Leinhardt 2010), without accounting for the
{\it gravitational} effect of the disk. Batygin \etal (2011) 
have considered disk 
gravity in the context of planet formation and evolution 
in systems with highly misaligned, distant ($10^2-10^3$ AU) 
stellar companions, affected by the 
Lidov-Kozai effect (Lidov 1961; Kozai 1962). However, this 
effect is probably irrelevant for planetesimal dynamics in 
small separation (tens of AU) binaries, which are likely 
coplanar with circumstellar disks.

In this Letter we show that apsidal precession of planetesimal 
orbits induced by disk gravity dominates secular evolution
of planetesimals at separations of several AU.
As a result, relative velocities at which bodies 
collide are reduced, sometimes by more than an order of 
magnitude. In massive disks this effect presents 
a natural solution of the fragmentation barrier issue
for the in-situ formation of the giant planets 
in small separation binaries, such as $\gamma$ Cephei.


\section{Secular evolution.}  
\label{sect:secular}


We consider planetesimal motion as Keplerian motion around
the primary perturbed by the gravity 
of the companion, that moves on larger, eccentric orbit, 
and the disk. Mass of the primary is $M_p$, and we define 
$\mu\equiv M_s/(M_p+M_s)$. We assume eccentricity of the 
stellar binary $e_b$ to be small and planetesimal orbit to be  
{\it coplanar} with the binary. 
Planetesimals are immersed in a massive, axisymmetric 
gaseous disk, characterized by surface 
density $\Sigma(r)$ specified in \S \ref{sect:disk}.

Assuming $e\ll 1$ the secular 
(averaged over the planetesimal and binary 
orbital motion) disturbing function for a planetesimal 
with semimajor axis $a$ and eccentricity vector
${\bf e}=(k,h)=(e\cos\varpi,e\sin\varpi)$ (with apsidal
angle $\varpi$ counted from the binary apsidal line, which 
is assumed fixed\footnote{Precession period of the secondary 
orbit exceeds important timescales of the problem.})
is (Murray \& Dermott 1999) 
\ba
R=na^2\times \left[\frac{1}{2}\left(A+\dot \varpi_d\right)
\left(h^2+k^2\right)-Bk \right],
\label{eq:R}
\ea
where
\ba
A &=& \frac{3}{4}\mu\frac{n_b^2}{n}\left(1+\frac{3}{2}e_b^2\right),
\label{eq:A}\\
B &=& \frac{15}{16}e_b \mu\frac{n_b^2}{n}\frac{a}{a_b}
\left(1+\frac{5}{2}e_b^2\right),
\label{eq:B}
\ea
and 
\ba
\dot \varpi_d=-\frac{1}{2n}
\left(\frac{2}{r}\frac{\partial U_d}{\partial r}+
\frac{\partial^2 U_d}{\partial r^2}\right)\Big|_{r=a}
\label{eq:disk_prec}
\ea
is the precession frequency of planetesimal orbit due to
the disk potential $U_d$. Here
$n_b=[G(M_p+M_s)/a_b^3]^{1/2}$ and $n=(GM_p/a^3)^{1/2}$ are 
the mean motions of the binary and planetesimal, respectively.
The contribution to $R$ proportional to $\dot \varpi_d$ 
arises from expansion of the disk potential along the 
eccentric planetesimal
orbit and averaging over its mean longitude.

Evolution equations for $h$ and $k$ are written using
$dh/dt=\left(na^2\right)^{-1}\partial R/\partial k$, 
$dk/dt=-\left(na^2\right)^{-1}\partial R/\partial h$ as
\ba
\frac{dh}{dt}=(A+\dot\varpi_d)k-B,~~~
\frac{dk}{dt}=-(A+\dot\varpi_d)h.
\label{eq:evol_eqs}
\ea
These equations agree with the work of Marzari \& Scholl 
(2000) as long as $\dot\varpi_d=0$. 

We write down the solution for 
${\bf e}(t)={\bf e}_{\rm free}(t)+{\bf e}_{\rm forced}(t)$,
where 
\ba
\left\{
\begin{array}{l}
k_{\rm free}(t)\\
h_{\rm free}(t)
\end{array}
\right\}
=e_{\rm free}
\left\{
\begin{array}{l}
\cos\left[(A+\dot\varpi_d)t+\varpi_0\right]\\
\sin\left[(A+\dot\varpi_d)t+\varpi_0\right]
\end{array}
\right\},
\label{eq:free_part}
\ea
$\varpi_0$ is a constant, and 
\ba
\left\{
\begin{array}{l}
k_{\rm forced}(t)\\
h_{\rm forced}(t)\\
\end{array}
\right\}
=e_{\rm forced}
\left\{
\begin{array}{l}
1 \\
0
\end{array}
\right\},
~~~e_{\rm forced}
=\frac{B}{A+\dot\varpi_d}.
\label{eq:forced_part}
\ea
Thus, free eccentricity vector ${\bf e}_{\rm free}$
rotates at a rate $A+\dot\varpi_d$ around the endpoint 
of the {\it fixed} forced eccentricity vector 
${\bf e}_{\rm forced}$. Note that setting $\dot\varpi_d=0$ 
we reproduce the solution obtained by Heppenheimer (1978).

Planetesimals starting on circular orbits have $e_{\rm free}=
e_{\rm forced}$ so that their eccentricity oscillates
with amplitude $2e_{\rm forced}$ and period 
$2\pi/(A+\dot\varpi_d)$.


\subsection{Disk model.}  
\label{sect:disk}

We model the disk as a constant $\dot M$ disk extending out
to the outer truncation radius $r_t$. Numerical simulations 
of accretion disks in binaries suggest that
$r_t\sim (0.2-0.4) a_b$ (Zhou \etal 2012), depending on $e_b$ 
and $\mu$. In our study we will commonly take $r_t=0.25 a_b$. 

Constant $\dot M$ assumption is a necessary simplification, 
which ignores the details of the disk structure at $r\sim r_t$. 
Assuming viscosity $\nu$ in the disk to be well described
by the radius-independent effective $\alpha$-parameter 
(Shakura \& Sunyaev 1973), one can write
\ba
\Sigma(r)=\frac{\dot M}{3\pi\nu}=\frac{\Omega\dot M}
{3\pi\alpha c_s^2}.
\label{eq:sig}
\ea
From this it is obvious that if the disk temperature scales
as $T\propto r^{-q}$ then the surface density behaves as
\ba
\Sigma(r)=\Sigma_0\left(\frac{r_0}{r}\right)^{p},~~~
p=\frac{3}{2}-q,
\label{eq:sig1}
\ea
where $r_0$ is some fiducial radius and $\Sigma_0=\Sigma(r_0)$.

Models of protoplanetary disks typically find $q$ to be close
to $1/2$, so that $p\approx 1$. In particular, the passive disk 
model of Chiang \& Goldreich (1997) has $q=3/7$, so that 
$p-1=1/14$. Outer parts of a disk in a binary are additionally 
heated by the radiation of the companion and tidal dissipation, 
so that $q$ may be lower than $3/7$ even for a passive
disk. For simplicity, in our calculations we will assume $p=1$,
which corresponds to a classical Mestel disk 
(Mestel 1963) if $r_t\to \infty$.

Assuming that the disk has power law profile (\ref{eq:sig1})
with $p=1$ all the way to $r_t$ we can express its surface 
density via the total disk mass $M_d$ enclosed within $r_t$ 
as
\ba
\Sigma(r)\approx \frac{M_d}{2\pi r_t}r^{-1}\approx 
2800~\mbox{g cm}^{-2}\frac{M_d}{10^{-2}M_\odot}
r_{t,5}^{-1} r_1^{-1},
\label{eq:sig3}
\ea
where $r_{t,5}\equiv r_t/(5$AU), 
and $r_{1}\equiv r/(1$AU). Interestingly, gas surface density 
at 1 AU in such a disk with $M_d=0.01M_\odot$ is not very 
different from that in a Minimum Mass Solar Nebula (MMSN; 
Hayashi 1981).


\subsection{Precession due to the disk.}  
\label{sect:precession}

A disk with the density profile (\ref{eq:sig1}) with $p=1$ 
extending to infinity is known to have constant circular 
velocity (Mestel 1963)
\ba
v_c=\left(r\frac{\partial U_d}{\partial r}\right)^{1/2}
=\left(2\pi G\Sigma_0 r_0\right)^{1/2}.
\label{eq:v_c}
\ea
Expressing $\partial U_d/\partial r$ from this relation
and plugging it in equation (\ref{eq:disk_prec}) we find
$\dot\varpi_d=-\pi G \Sigma(r)/(nr)$, where from now on we
use $r$ instead of $a$. Even though the circumprimary
disk in our problem is truncated at $r_t$ this expression
should still be able to give us a reasonable estimate of 
the precession rate $\dot\varpi_d$ due to the disk potential for 
$r\lesssim r_t$. Using equation (\ref{eq:sig3}) we find
\ba
\dot\varpi_d\approx -\frac{G M_d}{nr_t}r^{-2}=-n\frac{M_d}{M_p}
\frac{r}{r_t}.
\label{eq:omega_p}
\ea
Note that $\dot\varpi_d$ varies rather weakly with $r$, as $r^{-1/2}$,
which is consistent with Batygin \etal (2011).


\subsection{Planetesimal eccentricities.}  
\label{sect:ecc}

To assess the role of disk-driven precession on secular evolution
of planetesimals we evaluate
\ba
\frac{|\dot\varpi_d|}{A}\approx  \frac{4}{3}\frac{M_d}{M_s}
\frac{a_b^3}{r_t r^2}\approx 
20\frac{M_d/M_s}{10^{-2}}\frac{a_{b,20}^3}{r_{t,5}}r_1^{-2},
\label{eq:ratio}
\ea
where $a_{b,20}\equiv a_b/(20$AU). In making this estimate we 
neglected the term quadratic in $e_b$ in equation (\ref{eq:A}).
It is obvious that dynamics of planetesimals at several AU
is strongly affected by the disk-driven precession.
Indeed, $|\dot \varpi_d|$ exceeds $A$ for
\ba
r\lesssim r_{cr}\approx 4.6\mbox{AU}
\left(\frac{M_d/M_s}{10^{-2}}\frac{a_{b,20}^3}{r_{t,5}}
\right)^{1/2},
\label{eq:r_cr}
\ea
i.e. over essentially the whole assumed extent of the disk
even for the disk mass as small as $\sim 10^{-2}$ M$_\odot$.
Thus, if we are interested in planet formation at $2-3$ AU
we can neglect planetesimal precession due to the
secondary compared to the disk-driven precession, i.e.
neglect $A$ compared to $\dot\varpi_d$ in equation 
(\ref{eq:forced_part}) and other formulae.

Equation (\ref{eq:forced_part}) then predicts that the 
amplitude of eccentricity oscillations is 
\ba
e^{\rm disk}(r)&=& \frac{2B}{|\dot\varpi_d|}\approx 
\frac{15}{8}e_b\frac{M_s}{M_d}\frac{r^3 r_t}{a_b^4}
\label{eq:ampl}\\
&\approx & 3\times 10^{-3}\frac{e_b}{0.5}
\frac{0.01}{M_d/M_s}\frac{r_{t,5}}{a_{b,20}^4}r_1^3,
\ea
where we again neglected $e_b^2$ term in equation (\ref{eq:B}).
This is to be compared with 
\ba
e^{\rm n/disk}(r)=\frac{5}{2}\frac{r}{a_b}e_b\approx 
6.3\times 10^{-2}\frac{e_b}{0.5}\frac{r_1}{a_{b,20}},
\label{eq:ampl1}
\ea
which one finds neglecting disk-driven precession, i.e.
dropping $\dot\varpi_d$ in equation (\ref{eq:forced_part}).  
It is obvious that neglecting disk-driven precession leads
to an {\it overestimate} of the planetesimal eccentricity at
$\sim$AU separations by more than an order of magnitude.
This has important consequences for planetesimal growth as
we discuss further.


\subsection{Gas drag.}  
\label{sect:gas}

Equation (\ref{eq:ampl}) accounts for the presence of the disk
only through the precession caused by its gravity. However,
for small planetesimals the effect of gas drag is also important. 
Assuming quadratic drag force in the form 
${\bf F}\approx -v{\bf v}\rho_g/(\rho d)$ (here $d$ and $\rho$ 
are the object's radius and bulk density, $\rho_g\approx 
\Sigma/h$ is the gas density and $h$ is the disk scale height) 
we account for its effect on planetesimal dynamics 
by adding terms $-D\{h,k\}\left(h^2+k^2\right)^{1/2}$
to the first and second equations (\ref{eq:evol_eqs}), 
respectively (Marzari \& Scholl 2000). Here
\ba
D=n\frac{\rho_g r}{\rho d}=n\frac{\Sigma}{\rho d}\frac{r}{h}.
\label{eq:D}
\ea

For small planetesimal sizes (to be specified later by equation 
(\ref{eq:d_t})), in the gas drag-dominated regime, the drag 
force balances 
eccentricity excitation due to the secondary, i.e. the 
$B$ term in the first equation (\ref{eq:evol_eqs}). This results 
in the following estimate for the gas drag-mediated planetesimal 
eccentricity:
\ba
e_{gas}\approx \left(\frac{B}{D}\right)^{1/2}=
\left(e_b\frac{M_s}{M_p}\frac{h}{r}\frac{\rho d}{\Sigma}
\right)^{1/2}\left(\frac{r}{a_b}\right)^{2}.
\label{eq:e_gas}
\ea
This expression agrees with Paardekooper 
\etal (2008) and predicts that $e_{gas}\propto d^{1/2}$.
As a result, for small bodies one finds $e_{gas}<e^{\rm disk}$.

The transition between the drag-dominated behavior 
(\ref{eq:e_gas}) and the drag-free eccentricity scaling 
(\ref{eq:ampl}) occurs at the planetesimal size $d_{gas}$ 
where these two equations yield the same eccentricity:
\ba
d_{gas} & \approx & \frac{4n}{B}\frac{r}{h}\frac{\Sigma}{\rho}
e_{\rm forced}^2=
\frac{15}{8\pi}e_b\frac{r}{h}\frac{M_s}{M_d}
\frac{M_p r_t r}{\rho a_b^4}
\label{eq:d_t}
\\
& \approx & 1~\mbox{km}~\frac{e_b}{0.5}\frac{r/h}{30}
\frac{0.01}{M_d/M_s}\frac{M_{p,1}r_{t,5}r_1}
{\rho_3 a_{b,20}^4},
\ea
where $\rho_3\equiv\rho/(3$ g cm$^{-3})$, 
$M_{p,1}\equiv M_p/M_\odot$ and we have used 
equation (\ref{eq:sig3}). 

Planetesimal eccentricity behaves 
according to formula (\ref{eq:e_gas}) for $d\lesssim d_{gas}$,
and switches to drag-free regime (\ref{eq:ampl}) for 
$d\gtrsim d_{gas}$, see Figure \ref{fig:d_r}. 
The dependence of $d_{gas}$ on gas disk 
density and mass --- $d_{gas}\propto M_d^{-1}$ --- is somewhat 
counter-intuitive, since higher gas density results in 
stronger drag, making it more important for {\it larger} 
bodies. However, the gas-free planetesimal eccentricity 
(\ref{eq:ampl}) is itself a function of $M_d$ and decreases 
{\it faster} with increasing $M_d$ than does $e_{gas}$,
explaining the nontrivial $d_{gas}(M_d)$ dependence.


\section{Implications for planet formation.}  
\label{sect:planet}


Planetesimals grow in mutual collisions as long as their 
encounter velocity $v_{coll}$ (measured at infinity) 
is such that collisions do not result in the 
net loss of mass. The conditions for this depend, in 
particular, on planetesimal size and on whether planetesimals 
are strength- or gravity-dominated. Using results of 
Leinhardt \& Stewart (2012) for collisions of equal mass
(the most disruptive) strengthless bodies we roughly 
estimate the condition for planetesimal growth to be 
\ba
v_{coll}\lesssim 2v_{esc},
\label{eq:cond}
\ea
where 
the 
escape speed from the surface of an object of radius $d$ 
and bulk density $\rho$ is 
\ba
v_{esc}=\left(\frac{8\pi}{3}G\rho\right)^{1/2}d\approx
1.3~\mbox{m s}^{-1}\rho_3^{1/2}d_1
\label{eq:vesc}
\ea
(here $d_1\equiv d/(1$km)).
It becomes harder to break planetesimals when they are 
small enough for their internal strength to dominate 
over the gravitational energy, which is expected to 
happen for $d\lesssim d_s\sim 10$ km (Holsapple 1994). 

Planetesimal collisions occur at velocity of order 
$v_{coll}(r)\approx e^{\rm disk}v_K$, where $v_K=nr$ is the 
Keplerian speed. Using expression (\ref{eq:ampl})
we find
\ba
v_{coll}(r)\approx 90~\mbox{m s}^{-1}
\frac{e_b}{0.5}
\frac{0.01}{M_d/M_s}\frac{M_{p,1}^{1/2}r_{t,5}}{a_{b,20}^4}
r_1^{5/2}.
\label{eq:vcoll}
\ea
Plugging equations (\ref{eq:vesc}) and (\ref{eq:vcoll}) into 
the condition (\ref{eq:cond}) we find that erosion in equal-mass 
planetesimal collisions is avoided for bodies with 
$d\gtrsim d_{coll}$, where
\ba
d_{coll}\approx 35~\mbox{km}~\frac{e_b}{0.5}
\frac{0.01}{M_d/M_s}\left(\frac{M_{p,1}}
{\rho_3}\right)^{1/2}\frac{r_{t,5}}{a_{b,20}^4}
r_1^{5/2}.
\label{eq:d_coll}
\ea
Thus, for the fiducial 
binary parameters adopted here and for $M_d\sim 10^{-2}M_\odot$ 
only planetesimals larger than $\approx 35$ km would be able 
to grow at 1 AU. At 2 AU --- the semi-major axis of $\gamma$
Cephei Ab --- only bodies larger than $200$ km in radius 
would be able to survive in equal-mass collisions. 

However, in the absence of a disk the problem is much
worse: evaluating collisional velocity as $v_{coll}=
e^{\rm n/disk}v_K$ using equation (\ref{eq:ampl1}) and applying
condition (\ref{eq:cond}) one finds than in the absence of 
disk-induced precession only planetesimals larger than
\ba
d_{coll}^{\rm n/disk}\approx 700~\mbox{km}~\frac{e_b}{0.5}
\left(\frac{M_{p,1}}{\rho_3}\right)^{1/2}a_{b,20}^{-1}
r_1^{1/2}
\label{eq:d_coll1}
\ea
are able to survive in equal-mass collisions.
Clearly, collisional barrier appears {\it far} more severe 
if one disregards the effects of the disk on the secular 
evolution of planetesimals. We compare the behavior of 
$d_{gas}$, $d_{coll}$, and $d_{coll}^{\rm n/disk}$ 
as a function of $r$ in Figure \ref{fig:d_r}.

We also point out that $d_{coll}$ is very sensitive to the
binary semi-major axis $a_b$, unlike $d_{coll}^{\rm n/disk}$,
see equations (\ref{eq:d_coll}) and (\ref{eq:d_coll1}):
increasing $a_b$ from 20 AU to 30 AU reduces $d_{coll}$ by
a factor of $5$. The size of the region where disk-driven 
precession dominates secular evolution also expands rapidly
with increasing $a_b$, see equation (\ref{eq:r_cr}). To
summarize, properly accounting for the disk gravity 
considerably alleviates the collisional barrier in 
binaries, certainly for $r\lesssim 1$ AU.


\section{Planetesimal accretion is possible in massive disks}  
\label{sect:solution}

We now propose a solution to the problem of planetesimal 
accumulation in binaries, raised in \S \ref{sect:intro}. 
We argue that if 
\begin{itemize} 
\item disk is {\it massive}, $M_d\sim 0.1M_\odot$, 
\item planetesimals are {\it strength-dominated below 
$\sim 10$} km, 
\end{itemize}
then the fragmentation barrier can be overcome even 
at separations of $\approx 2$ AU, where planets in several 
binaries are found.

\begin{figure}
\plotone{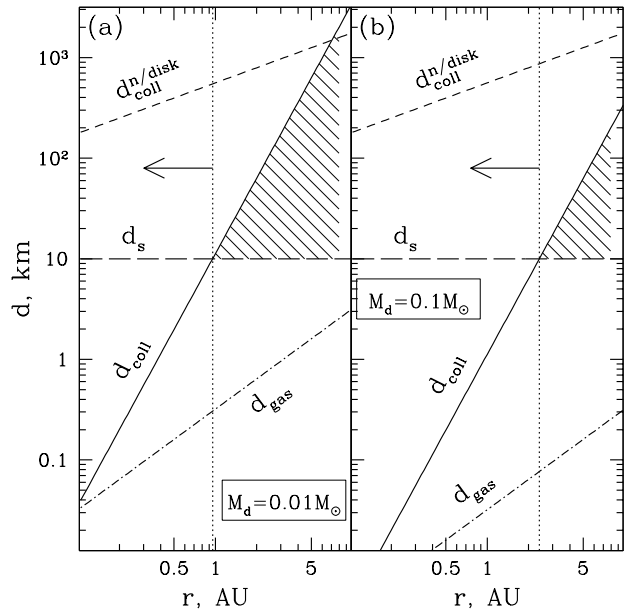}
\caption{
Characteristic planetesimal sizes vs. radius for two disk 
masses $M_d$: (a) $0.01M_\odot$ and (b) $0.1M_\odot$. We 
display $d_{coll}$ ({\it solid}, eq. [\ref{eq:d_coll}]), 
$d_{gas}$ ({\it dot-dashed}, eq. [\ref{eq:d_t}]),
$d_{coll}^{\rm n/disk}$ ({\it short-dashed}, eq. 
[\ref{eq:d_coll1}]), and planetesimal radius $d_s=10$ km 
below which we consider objects as strength-dominated 
({\it long-dashed}). The two latter sizes do not depend 
on $M_d$. Calculations are done for $e_b=0.4$, 
$M_s=0.4M_\odot$, $a_b=20$ AU, $M_p=M_\odot$ 
(typical for small separation binaries,
Chauvin \etal 2011), $r_t=5$ AU, and $r/h=30$.
Planetesimals
in the shaded region (``danger zone'') get destroyed in
equal-mass collisions according to criterion (\ref{eq:cond})
precluding planetary growth at corresponding separations.
Accretion-friendly zone is to the left of the vertical 
dotted line in each plot; it is wider for higher $M_d$ and
extends to $\approx 2.5$ AU for $M_d=0.1M_\odot$. 
\label{fig:d_r}}
\end{figure}

Equation (\ref{eq:d_coll}) shows that higher $M_d$ results in
smaller planetesimal size $d_{coll}$, below which strengthless 
objects are destroyed or eroded in equal-mass collisions. 
Smaller planetesimals are (1) more resistive to collisional 
erosion because of their internal strength and (2) stronger 
affected by gas drag. When the latter dominates, planetesimal 
velocities are reduced and collisions are less destructive. 
However, increasing $M_d$ reduces not only $d_{coll}$ but
also $d_{gas}$ in such a way that $d_{coll}/d_{gas}$ stays
constant. At 1 AU this ratio is about 30 so that independent
of $M_d$ there is still a significant ``danger zone'' between 
the planetesimal size $d_{gas}$ below which gas drag lowers 
$v_{coll}$ helping accretion and the radius $d_{coll}$ above 
which colliding strengthless bodies can grow, see Figure 
\ref{fig:d_r}.

On the other hand, equation (\ref{eq:d_coll1}) predicts that 
$d_{coll}\lesssim d_s$ at $r=2$ AU for $M_d/M_s=0.2$, 
if internal strength dominates over the gravitational binding 
energy of the body with $d_s=10$ km making 
it harder to erode or destroy. This is a resolution of the 
collisional barrier problem in binaries that we favor in 
this work. 

\begin{figure}
\plotone{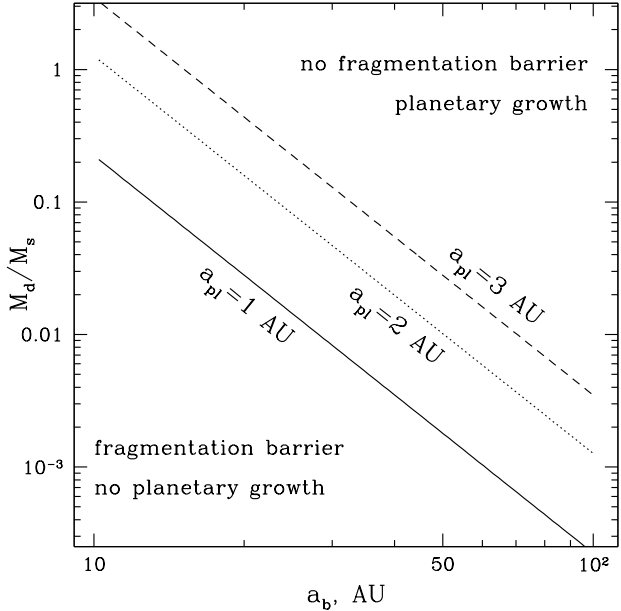}
\caption{
Plot of $a_b-M_d/M_s$ phase space illustrating conditions under
which planets can form in binaries, assuming $d_s=10$ km. 
Different curves show the 
relation (\ref{eq:phase}) for different values of the planetary 
semi-major axis $a_{pl}$, interior to which planet formation is 
possible: $a_{pl}=1$ AU ({\it solid}), $a_{pl}=2$ AU ({\it dotted}), 
$a_{pl}=3$ AU ({\it dashed}). Calculation assumes $e_b=0.4$,
fixed $r_t/a_b=0.25$, $a_b=20$ AU, and $M_p=M_\odot$. At a 
given $a_{pl}$ fragmentation barrier 
is avoided and planet formation proceeds smoothly through the 
planetesimal stage to the right (and above) of 
the corresponding line. 
\label{fig:phase}}
\end{figure}

To avoid fragmentation barrier we thus require that 
$d_{coll}<d_s$. Using equation (\ref{eq:d_coll}) we can 
rephrase this condition in the form of a lower limit on 
the disk mass at a given separation $a_{pl}$
from the primary:
\ba
\frac{M_d}{M_s}\gtrsim 0.035~\frac{e_b}{0.5}
\left(\frac{M_{p,1}}{\rho_3}\right)^{1/2}\frac{r_t/a_b}{0.25}
a_{pl}^{5/2}a_{b,20}^{-3}.
\label{eq:phase}
\ea
In Figure \ref{fig:phase} we illustrate this constraint as a 
function of the binary semi-major axis, for different values 
of $a_{pl}$. It is clear that in {\it very small} separation 
binaries with $a_b=10$ AU growing planets even at 1 AU 
requires a massive disk, $M_p\approx 0.2M_s$.

External companions in giant planet-hosting binaries typically 
have mass $M_s\approx 0.4M_\odot$ (Chauvin \etal 2011), meaning 
that our scenario of planet formation at 2 AU needs 
$M_d\approx 0.1M_\odot$. Such disk mass  
may seem high but it is also the case that planet-hosting 
binary systems with $a_b\approx 20$ AU that we consider 
here contain more mass in total than the 
descendants of the typical T Tauri stars. 

One might worry that such massive disks would be prone to 
gravitational instability (GI). With the density profile 
(\ref{eq:sig3}) we estimate the Toomre 
$Q\equiv n c_s/(\pi G \Sigma)$ ($c_s$ is the sound 
speed) as
\ba
Q\approx 2\frac{M_p}{M_d}
\frac{h}{r}\frac{r_t}{r}\approx 3\left(\frac{0.1}{M_d/M_p}\right)
\left(\frac{30}{r/h}\right)\frac{r_{t,5}}{r_1}.
\label{eq:Q}
\ea
Thus, even for 
$M_d=0.1M_p\approx 0.1M_\odot$ the disk is at most marginally 
unstable to GI at $2$ AU. However, even if it were unstable, the 
surface density and 
optical depth at this distance would be so high that the cooling
time would far exceed the local dynamical time, making planet
formation by direct disk fragmentation impossible 
(Gammie 2001; Rafikov 2005). Instead, the disk would slowly 
evolve under the action of gravitoturbulence (Rafikov 2009).

On the other hand, high $M_d$ simplifies planet formation 
in other ways. In particular,
planets in these systems are quite massive (Chauvin \etal 2011)
and larger $M_d$ provides mass reservoir for their assembly. 
Higher surface density of the protoplanetary 
disk also means larger isolation mass (Lissauer 1993) possibly making
it high enough at 2 AU to trigger core accretion without 
the need to go through the long-lasting stage of giant 
impacts (Chambers 2004). Higher $\Sigma$ likely implies larger
dead zone (Gammie 1996) in the disk, providing quiet 
conditions for planetesimal formation and growth, and 
resulting in smaller viscosity, which, possibly, means longer 
disk lifetime. The timescale on which planets form also goes 
down as $M_d$ increases.

Another potential solution to 
the collisional barrier problem in binaries is the direct 
formation of large planetesimals by e.g. streaming and/or 
gravitational instabilities  (Johansen \etal 2007; 
Th\'ebault 2011). Large $M_d$ and small $d_{coll}$ are 
helpful for this mechanism as well since to overcome 
fragmentation barrier in a massive disk such instabilities 
would only need to produce bodies with sizes of tens of km, 
rather than $\sim 10^3$ km dwarf planets.


\section{Discussion.}  
\label{sect:summ}


We now mention several additional factors that may strengthen or 
weaken our conclusions. First, our collisional growth condition 
(\ref{eq:cond}) may be too stringent. Previously, using
a more refined fragmentation criterion Th\'ebault (2011)
found that in HD 196885 ($M_p=1.3M_\odot$, $a_b=21$ AU$, 
e_b=0.42$) planetesimal growth is possible even in the absence 
of disk-driven precession at $2.6$ AU as long as the planetesimal 
size exceeds 250 km. However, according to our formula 
(\ref{eq:d_coll1}) with the same assumptions (same system 
parameters and $\dot \varpi_d=0$), growth is possible
only for $d\gtrsim 10^3$ km. Thus, our fragmentation 
criterion likely {\it overestimates} planetesimal size 
above which objects grow efficiently, and, in fact, it might 
be {\it easier} to overcome the fragmentation barrier with 
the more realistic growth condition than our simple 
criterion (\ref{eq:cond}). At a given distance this
would {\it lower} the value of $M_d$ needed to overcome
fragmentation barrier. 

Second, even if planetesimals are collisionally weak their 
growth may still proceed mainly via unequal-mass collisions 
(which more frequently result in mergers) if
the number of relatively massive objects is small 
(Th\'ebault 2011). Outward migration of planets by 
scattering of planetesimals has also been 
invoked (Payne \etal 2009) to explain planets on
AU-scale orbits in small separation binaries.

On the other hand, there are also factors complicating 
planetesimal growth. In particular, eccentricity of the 
gaseous disk induced by the companion may affect 
planetary growth at small sizes ($d\lesssim d_{gas}$). 
Also, gas drag-induced inspiral of planetesimals may
deplete the disk of some solids. The relative importance 
of these factors for planet formation in binaries will 
be assessed in the future.


\acknowledgements

This work was supported by NSF via grant AST-0908269.


\end{document}